\documentclass{article}

\usepackage{PRIMEarxiv}

\usepackage[utf8]{inputenc} % allow utf-8 input
\usepackage[T1]{fontenc}    % use 8-bit T1 fonts
\usepackage{hyperref}       % hyperlinks
\usepackage{url}            % simple URL typesetting
\usepackage{booktabs}       % professional-quality tables
\usepackage{amsfonts}       % blackboard math symbols
\usepackage{nicefrac}       % compact symbols for 1/2, etc.
\usepackage{microtype}      % microtypography
\usepackage{lipsum}
\usepackage{fancyhdr}       % header
\usepackage{graphicx}       % graphics
\graphicspath{{media/}}     % organize your images and other figures under media/ folder
\usepackage{bm}
\usepackage{subfigure}

\newcommand{\mf}{\mathbf}

\newenvironment{eq}{\begin{equation}}{\end{equation}}

  \title{Audible networks: deconstructing and manipulating sounds with Deep Non-Negative Autoencoders

}

\author{
  Juan José Burred \\
  Department of Music, CNMAT \\
  University of California, Berkeley \\
  Berkeley, CA, US\\
  \texttt{jjburred@jjburred.com} \\
   \And
  Carmine-Emanuele Cella \\
  Department of Music, CNMAT \\
  University of California, Berkeley \\
  Berkeley, CA, US\\
  \texttt{carmine.cella@berkeley.edu} \\
}

\begin{document}
\maketitle

\begin{abstract}

We propose the use of Non-Negative Autoencoders (NAEs) for sound deconstruction and user-guided manipulation of sounds for creative purposes. NAEs offer a versatile and scalable extension of traditional Non-Negative Matrix Factorization (NMF)-based approaches for interpretable audio decomposition. By enforcing non-negativity constraints through projected gradient descent, we obtain decompositions where internal weights and activations can be directly interpreted as spectral shapes and temporal envelopes, and where components can themselves be listened to as individual sound events. In particular, multi-layer Deep NAE architectures enable hierarchical representations with an adjustable level of granularity, allowing sounds to be deconstructed at multiple levels of abstraction—from high-level note envelopes down to fine-grained spectral details. This framework enables a wide new range of expressive, controllable, and randomized sound transformations. We introduce novel manipulation operations including cross-component and cross-layer synthesis, hierarchical deconstructions, and several randomization strategies that control timbre and event density. Through visualizations and resynthesis of practical examples, we demonstrate how NAEs can serve as flexible and interpretable tools for object-based sound editing.

\end{abstract}

% keywords can be removed
\keywords{Non-Negative Autoencoders \and Sound deconstruction \and Neural sound synthesis \and Unsupervised Source Separation \and Interpretable AI}

\section{Introduction}

In a broad sense, Machine Learning methods seek to detect and extract underlying patterns from data. These learned patterns are expressed as mathematical models that can be used
for automatic classification tasks (e.g., computer vision, speech recognition, music transcription), for semantically meaningful transformations (e.g., editing objects from images, separating
instruments from musical mixtures) or for the automatic generation of new data (e.g., image and speech synthesis, generative music). This latter category in particular (generative AI) has seen staggering progress in recent years, and has
led to fully or highly automated systems capable of creating realistic images, videos, or music with very little user intervention. 

Our research focuses on ML systems that, in contrast to the above-mentioned generative methods, rely on a fair amount of user interaction and collaboration, and that allow a close expressive control of the results, thus keeping composers and performers fully engaged in the creative process. More specifically, we aim at leveraging the analytic power of ML algorithms to create collections of sonic elements,
i.e., sets of elementary but relatively structured sounds that the composer or performer can subsequently freely combine, mix, and manipulate. 

Conventional sound synthesis and modification tools are based on manipulating elements of a low
semantic level, such as individual samples, instantaneous frequencies, amplitudes, global filter shapes or envelopes,
etc. Typically, an input sound is converted into a time-frequency representation, such as a spectrogram, which is
then modified and inverted to obtain the new sound. In computer music, this is called the \textit{analysis/resynthesis paradigm}. % (**CITE ROADS?**).
In contrast, when applied to sounds, ML algorithms can infer elements of a higher semantic level, such as full musical notes, rhythmical motifs, phonemes, vibration patterns, etc. Furthermore, they also have the ability to reveal hidden structures that are less obvious to hear from the initial sound. Ultimately, we strive towards developing a new paradigm for \textit{object-based} sound editing. To emphasize the fact that the goal is to obtain objects of a rich spectral and temporal structure, we call this process \textit{sound deconstruction} rather than analysis or decomposition. 

The domain of ML that lends itself naturally to this goal is \textit{source separation}, which can be
understood as a subfield of ML where the recognized patterns are expected to overlap each other in the observed
space. Indeed, it is almost certain that a constituent sonic event will have some degree of frequency and time overlap with the rest of the sound. When added back together, the elements obtained via source separation should reconstruct the original sound as closely as possible. In other words, the obtained sonic elements are in fact layers that make up the analyzed sounds when mixed
back together.

Another key requirement of this line of research is that the source separation process must be fully \textit{unsupervised}, meaning that the analyzed dataset is not pre-labeled into any categories. The algorithm must be capable of discovering by itself salient and interesting structural components of any kind of sound.
This contrasts with supervised source separation, where the goal is to separate a mixture of sounds into a pre-defined
set of sources, such as when unmixing a song into a given and immutable set of instruments \cite{demucs}.

Before the explosion of Deep Neural Networks (DNN) around 2015, the predominant ML method for sound source separation was Non-Negative Matrix Factorization (NMF) \cite{lee99}. In its simplest form, NMF is an unsupervised pattern recognition method that, when applied to a time-frequency representation such as a power or magnitude spectrogram, delivers a set of spectral and temporal shapes contained in the sound, named respectively \textit{bases} and \textit{activations}. The components that NMF extracts are not only interpretable: they are sonic objects that can be actually listened to. 

Since then, as in virtually all other domains of ML and AI, DNNs are the unquestionably best-performing algorithms for sound source separation. They excel in supervised separation tasks, for which they are trained on large labeled datasets of perfectly separated sources. On the other hand, and in contrast to NMF, the internals of neural networks, let alone deep ones, are extremely hard to interpret. There are no easily recognizable spectral or temporal shapes when observing how most DNNs internally process sounds, and therefore it is hard to predict how manipulating the learned model will affect the output. This is not a problem for supervised separation tasks, where the goal is to obtain the best objective separation quality possible in a fully automatic way. However, interpretability is essential to our task at hand, and therefore most DNN architectures are not appropriate for object-level sound manipulation.

\begin{figure}[b!]
  \centering
  \includegraphics[width=17cm]{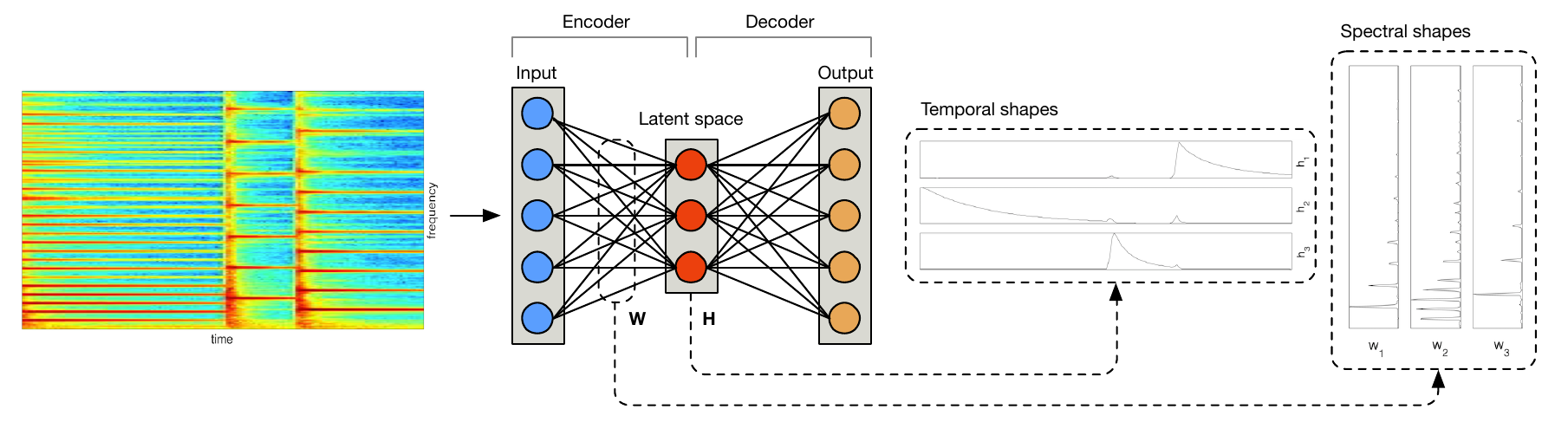}
  \caption{Conceptual representation of a Non-negative Autoencoder (NAE) for sound analysis. An input spectrogram (left) is fed to the network. The weights ($\mf{W}$) and activations ($\mf{H}$) of the bottleneck layer can be interpreted as a set of, respectively, temporal and spectral elements of the input sound.}
  \label{fig:nae}
\end{figure}

Still, DNNs possess many qualities that seem desirable for user-guided sound manipulation: flexibility, scalability, the ability to handle large datasets, and a hierarchical architecture that allows working at different levels of abstraction. In the context of source separation, the reformulation of NMF as a neural network was proposed in 2017 \cite{smaragdis17}. In that work, the non-negativity constraint that lies at the core of NMF was applied to an autoencoder, which is a type of neural network with an inner layer, called "bottleneck", with much fewer neurons than the input and output layers. Minimizing a reconstruction error between the output and the input during training forces the autoencoder to learn an efficient representation of the sound at the bottleneck layer, called the \textit{code} or the \textit{latent representation}. The non-negativity constraint simply means that all elements in all involved matrices (input, output, and inner weights and activations) are forced to be zero or positive. Imposing this on an autoencoder makes the internal weights and activations interpretable as, respectively, spectral and temporal shapes, in the same way as NMF, thus obtaining a Non-Negative Autoencoder (NAE). Fig. \ref{fig:nae} shows an example of a shallow (1-inner-layer) NAE, where the input amplitude spectrogram has been decomposed as a set of temporal shapes and spectra. A temporal shape can be interpreted as a time-varying signal that modulates the amplitude of its corresponding spectral shape. When all modulated spectra are summed, a close approximation to the input spectrogram is obtained. 

Previous sound-related applications of NAEs have been focused on source separation \cite{smaragdis17, suarez20, zunner21, ozer22}. In this work, we propose the use of NAEs, both shallow and deep, as new tools for sound design and manipulation. By extending the structured deconstruction capabilities of NMF into a multi-layer, hierarchical representation, we add abstraction level as a new controllable dimension. This allows a range of new sound manipulation and randomization operations, which are incrementally fine-grained. We will discuss the challenges associated with obtaining a multi-layer and interpretable representation that is useful as both a framework for sound analysis and an expressive tool for sound creation and performance.

\section{Related work}

There exists a body of previous research work dedicated to the unsupervised decomposition of sounds, but mostly applied to speech enhancement, denoising, and sound event detection \cite{wisdom20, Schulze-Forster2022}. Very little research has been previously dedicated to the computational extraction of elementary sounds for resynthesis and creation. Conceptually related are \textit{granular synthesis}  \cite{roads04} and \textit{concatenative synthesis}\cite{schwarz05}, but in those cases the elements are very short, non-overlapping segments without temporal structure. There also exist approaches to \textit{neural synthesis} \cite{rave, nsynth}, but without the unsupervised decomposition idea. A more relevant work is the \textit{theory of sound types} \cite{cella11, cella13}, where ML approaches are used to derive a set of prototypical sound segments, albeit still without temporal structure. In terms of the ability to manipulate the internals of neural networks for creative purposes, a related concept is \textit{network bending} \cite{broad21, dzwonczyk24}.

Our own previous research \cite{burred13, burred14} focused on using NMF as a front-end for sound deconstruction and manipulation. This led to the development of Factorsynth \cite{burred16}, a software tool that acts as a graphical interface to edit and recombine the components issued from spectrogram factorizations. The present work constitutes the next natural step of this line of research, where we replace NMF with multi-layer neural architectures for increased versatility. 

The introduction of non-negative constraints as a way to obtain interpretable neural networks was first proposed in \cite{chorowski15} in the context of image recognition. Direct reformulations of NMF as a neural network were then proposed in \cite{flenner17} for text and audio classification and in \cite{smaragdis17} for sound source separation. Subsequent work focused on investigating shallow NAEs in terms of convergence behavior, optimization algorithms, and computational performance \cite{suarez20, zunner21, ozer22}. Those works confirm the functional equivalence between shallow NAEs and NMF. In contrast to shallow NAEs, deep NAEs (DNAEs) of more than one inner layer have been seldom explored in the audio domain.

\section{Non-Negative Autoencoders (NAEs)}

NMF approximates an input non-negative matrix $\mf{X}$ as a product of two non-negative matrices: $\mf{X} \approx \mf{WH} = \hat{\mf{X}}$. When $\mf{X}$ is a real-valued, non-negative time-frequency representation, such as an amplitude or power  spectrogram of size $F \times T$ ($F$ frequency bins and $T$ time frames), the $K$ columns of $\mf{W}$ (called \textit{bases)} can be interpreted as a set of spectra, and the $K$ rows of $\mf{H}$ (called \textit{activations}) can be interpreted as a set of temporal functions that modulate their corresponding spectra from matrix $\mf{W}$. Matrix $\mf{W}$ is of size $F \times K$, and matrix $\mf{H}$ is of size $K \times T$, where $K$ (the number of spectra/activations) is a parameter set by the user. The shallow NAE illustration of Fig. \ref{fig:nae} contains a representation of the same type of bases and activations NMF would produce. To obtain such a decomposition, a certain loss function $D(\mf{X},\hat{\mf{X}})$
 is defined to measure the error between the input and the approximation, often a Kullback-Leibler divergence, which is then minimized using gradient descent. The most popular way to enforce non-negativity during NMF optimization is to modify the gradient descent updates in such a way that they only contain matrix multiplications, since multiplying zero or positive numbers together can never result in negative values. These kinds of optimization rules are called \textit{multiplicative updates} \cite{lee99}.

A natural neural reformulation of this decomposition requires a network with an input layer of $F$ units (one per frequency bin), an inner bottleneck layer of $K \ll F$ units, and an output layer of also $F$ units. Each spectral frame is considered a separate sample fed to the network, and thus $T$ is the total number of samples. To imitate NMF, the whole input spectrogram $\mf{X}$ is typically passed as a single batch, and we follow the same convention here. Furthermore, the bias vectors are set to zero. We thereby obtain a shallow autoencoder defined as follows:
\begin{eqnarray}
	\mathrm{Encoder}: & \mf{H}  =  g ( \mf{W}_e \mf{X}) \label{eq:shallow_encoder} \\ 
	\mathrm{Decoder}: & \hat{\mf{X}}  =  g ( \mf{W}_d \mf{H}) \label{eq:shallow_decoder}
\end{eqnarray}
The encoder is a fully-connected layer that reduces the dimensionality of the input $\mf{X}$ by multiplying it by a weight matrix $\mf{W}_e$ and passing the result through an activation function $g(\cdot)$. The resulting latent representation $\mf{H}$ ($K$ units, $T$ samples) is equivalent to the NMF activation matrix $\mf{H}$. The decoder increases the dimensionality via the inverse operation via a decoder weight matrix $\mf{W}_d$, which is functionally equivalent to NMF's $\mf{W}$ matrix. The encoder weight matrix $\mf{W}_e$ is a new matrix without a counterpart in NMF, and is required in order to simulate the reduction to a low-dimension using a neural architecture.

We still have to impose non-negativity on all elements of all matrices involved in the autoencoder, in order for it to be internally interpretable. Imposing non-negativity on $\mf{H}$ is straightforward by using an activation function which only produces non-negative outputs, such as ReLU ($g(x) = \max(0,x)$) or Softplus ($g(x) = \ln (1 + e ^ x)$). 

Enforcing non-negativity on the weight matrices $\mf{W}_e$ and $\mf{W}_d$ is less obvious, and several strategies have been proposed to that end. In \cite{smaragdis17}, the non-negativity of the weights is indirectly encouraged by adding a sparsity regularizer on $\mf{H}$ to the loss function. The underlying idea is that maximizing the sparsity of the latent representation results in less redundant weight matrices, with fewer elements that need to cancel out to improve the representation. This has been shown to improve the percentage of non-negative elements; however, it cannot guarantee 100\% non-negativity. Furthermore, a sparse $\mf{H}$ prevents efficiently representing sounds comprised of components overlapping in time. This approach was therefore discarded for the present work.

An alternative approach was proposed in \cite{zunner21} and \cite{ozer22} by deriving a set of multiplicative update rules for a shallow NAE, similar to the ones for NMF, and thus ensuring non-negativity during optimization. However, this approach is not scalable, since a new set of rules would have to be derived for the addition of each new layer to the architecture, making it impractical for deep NAEs, which we study here.

Instead, we use a \textit{projected gradient descent} method \cite{suarez20, lin07} in which the negative weights are rectified to zero after every optimization step, and crucially also right after the initial random initialization of the matrices. Even though we are only focusing on the interpretability of the decoder weight matrix $\mf{W}_d$ in the current study, we have observed much better training stability when applying the gradient projection constraint on both $\mf{W}_e$ and $\mf{W}_d$.

Given the shallow NAE of Eqs. \ref{eq:shallow_encoder} and \ref{eq:shallow_decoder}, we can extend it to an $L$-layer Deep NAE (DNAE) as follows:
\begin{eqnarray}
	\mathrm{Encoder}: & \mf{H}  & =  g(\mf{W}_{e1} \ldots g(\mf{W}_{e(L-1)} \cdot  g ( \mf{W}_{eL} \mf{X})) ) \\ 
	\mathrm{Decoder}: & \hat{\mf{X}} & =  g(\mf{W}_{dL}   \ldots  g(\mf{W}_{d2} \cdot g ( \mf{W}_{d1} \mf{H}) ))
\end{eqnarray}
For simplicity, we have assumed the same non-linearity $g(\cdot)$ in all layers, and no bias vectors. An implicit assumption for this deep definition is that the number of units gradually decreases throughout the layers of the encoder, and increases symmetrically throughout the layers of the decoder. Note that under this convention, an $L$-layer DNAE is actually a network of $2L+1$ layers: the $L$ encoder layers, the $L$ symmetrically arranged decoder layers, plus the innermost latent layer.

\section{Sound deconstruction with NAEs}

Our first range of experiments consisted of visualizing sound deconstructions of simple mono mixtures with shallow and deep NAE variants. We created multi-layer displays based on the decoder part of the NAE in order to examine the resulting internal bases and activations, and used these representations to evaluate a range of optimization algorithms and loss functions. A configuration that consistently resulted in stable optimizations and easily explainable components was using RMSprop as the optimization algorithm, Glorot initialization (rectifying the negative elements) to initialize all weight matrices, and Generalized Kullback-Leibler divergence as the loss function, defined as:
\begin{eq}
\mathcal{L} =  D_{GKL}({\mathbf{X}}, \hat{\mathbf{X}}) = \frac{1}{FT} \sum_{f=1}^{F} \sum_{t=1}^{T} \left[ \mathbf{X}_{(f,t)} \left( \log \mathbf{X}_{(f,t)} - \log \hat{\mathbf{X}}_{(f,t)} \right) - \mathbf{X}_{(f,t)} + \hat{\mathbf{X}}_{(f,t)} \right]\label{eq:gkl}
\end{eq}

Using the average as the reduce operator in Eq. \ref{eq:gkl}, instead of the more common addition, helped avoiding gradient explosion during training by keeping the magnitude of the loss and its derivatives bounded across varying input sizes. We used ReLU as nonlinearity for the inner layers, and Softplus as the non-linearity of the output layer, which results in softer spectrograms.

It is worth mentioning that this use of a neural network differs significantly from the traditional approach of analyzing a large dataset of sounds to create generalized models, as required, for example, in generative applications or supervised source separation. In our case, we analyze a single sound file—albeit possibly long and complex. The dataset consists on the collection of individual frames from the input spectrogram, and the entire file is passed as a single batch during optimization. Overfitting is not a concern; on the contrary, it is desirable, as we are not aiming to generalize to unseen sounds, but rather to obtain the best possible representation of the sound under study.

We will illustrate the type of visualizations obtained using a toy mixture of two harmonic sounds (one consisting of 3 consecutive trumpet notes of the same pitch, the other consisting of 2 consecutive trumpet notes of a lower pitch) and a wideband sound consisting of 3 consecutive bursts of white noise. More visualizations and sound examples can be found on the companion website\footnote{\url{https://jjburred.com/research/audible2025}}. 

\begin{figure}[t!]
  \centering
  \includegraphics[width=17cm]{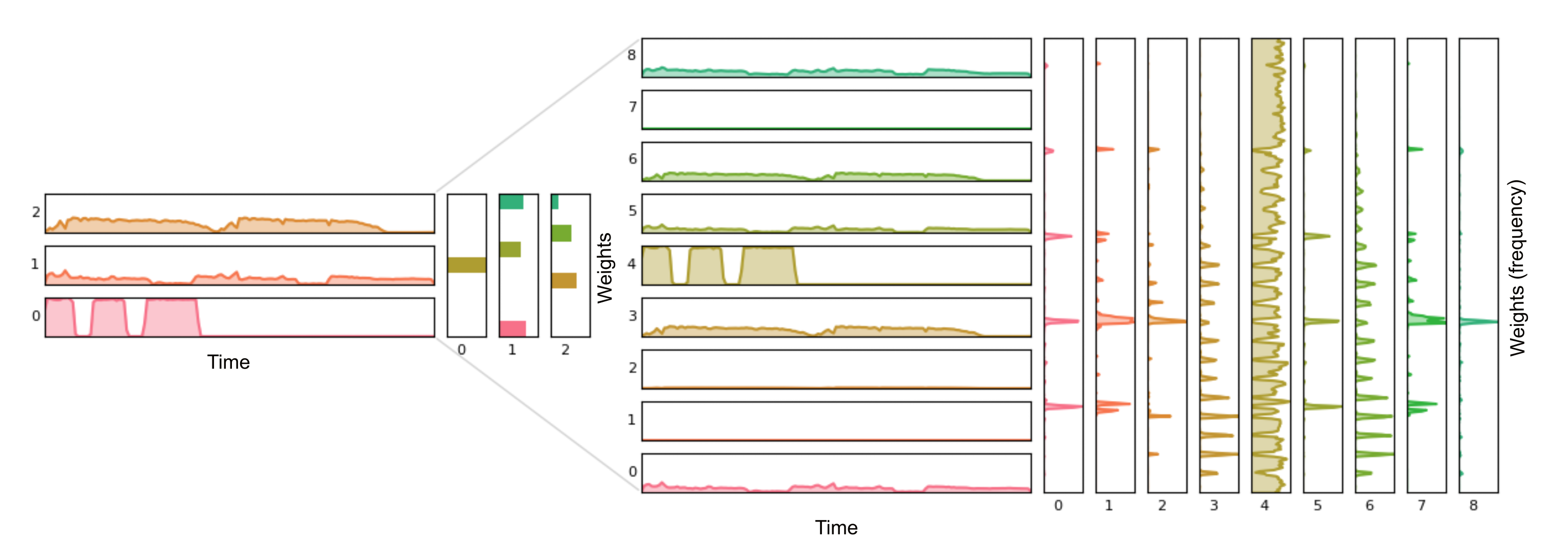}
  \caption{Visualization of a 2-layer NAE deconstruction of a simple mixture of 3 sources.}
  \label{fig:nae_2_layer}
\end{figure}

Figure \ref{fig:nae_2_layer} shows the visualized deconstruction of the described sound into a 2-layer NAE (5 layers in total), with 3 units in the latent layer and 9 in the outer layer. The horizontal functions are the extracted temporal shapes, corresponding to the output of the neuron activations. The distinct amplitude envelopes of the 3 sources of the described mixture sound (2 notes, 3 notes, and 3 square bursts) are clearly visible as the activations of the inner layer. The vertical plots are the learned weights of the network. In the case of the outer layer, the weight vectors (each of size $F$) are directly interpretable as spectra, with low frequencies at the bottom and high frequencies at the top of the figure. The weights of the inner layer are plotted as bar graphs, and can be understood as intermediate mixing factors. They determine how much of each activation of the previous layer will be present in the next layer, and how they are distributed among units as their number increases towards the output. For instance, the first weight vector of the inner layer (labeled as 0) contains a single bar, which means that the corresponding inner activation (the amplitude envelope of the 3 square noise bursts) propagates towards a single unit (number 4) of the outer layer. As a consequence, it can be seen that spectrum number 4 is the only one that contains wideband noise. The colors help in following the propagation across layers. 

% model 2025-07-08_23_13_19.008407

While some outer spectra are redundant (3 and 6, 0 and 5), others exhibit distinct features even when associated with the same inner activation, such as spectra 0 and 1. In this example, there is also an instance of a component that is shared between two of the sources: the single-partial spectrum number 8. The fact that this is a shared component can be easily noticed by inspecting the inner weight factors: the upper weight bar (depicted here in green) is present in both components 1 and 2, meaning those two sources have both contributed to the same output spectrum. These are simple instances of the kind of detailed analyses enabled by the extensible NAE architecture. 

\subsection{Resynthesis}

Besides visual inspection, we want to be able to listen to the generated components, both individually and in any combination. When performing the outer product between a given temporal shape of the outer layer $\mf{h}_{(k,L)}$ and a given spectrum of the outer layer $\mf{w}_{(k,L)}$, the result is a magnitude spectrogram $\mf{C}_{(k,L)} = \mf{w}_{(k,L)} \otimes  \mf{h}_{(k,L)}$ of size $F \times T$, representing the time-frequency evolution of that particular activation/base combination. In order to resynthesize that particular component, it is necessary to generate or attach the missing phase information and invert the spectrogram. As in \cite{burred14, smaragdis17}, we compute a Wiener mask for that component, to which we attach the phase matrix $\mf{\Phi}$ of the input sound:
\begin{eq}
	\mf{S}_k = \mf{M}_k \circ \mf{X} \circ e^{j \mf{\Phi}},
\end{eq}  
where $\circ$ denotes element-wise multiplication and $\mf{M}_k$ is a unity-bounded time-frequency mask given by
\begin{eq}
	\mf{M}_k = \frac{\mf{w}_{(k,L)} \otimes  \mf{h}_{(k,L)}}{\sum_{k=1}^{K_L} \mf{w}_{(k,L)} \otimes  \mf{h}_{(k,L)}}. \label{eq:wiener}
\end{eq} 
The result is a complex-valued Short Time Fourier Transform (STFT) matrix $\mf{S}_k$ that can then be inverted into the time domain. This definition of the mask ensures that all components add up exactly to the initial mixture, a condition named the \textit{conservativity constraint}.

\subsection{Deep deconstruction}

When increasing the number of decoder layers to 3 or more (total layers 7 or more), we have observed that it is necessary to impose additional constraints on the GKL loss of Eq. \ref{eq:gkl} in order to keep diversity in the extracted components and limit redundance in the representations. To this end, we introduce a sparsity regularization term to the loss function of \ref{eq:gkl}, acting on the outer weights:
\begin{eq}
	\mathcal{L}_s = D_{GKL}({\mathbf{X}}, \hat{\mathbf{X}}) + \lambda ( \| \mf{W}_{eL} \| + \| \mf{W}_{dL} \| ).
\end{eq}
Even though the spectra used for visualization and resynthesis stem from the outer decoder weight matrix $\mf{W}_{dL}$, we have observed it to be more effective to apply the sparsity constraint to both the encoder and decoder outer weight matrices. 

\begin{figure}[t!]
  \centering
  \includegraphics[width=16.5cm]{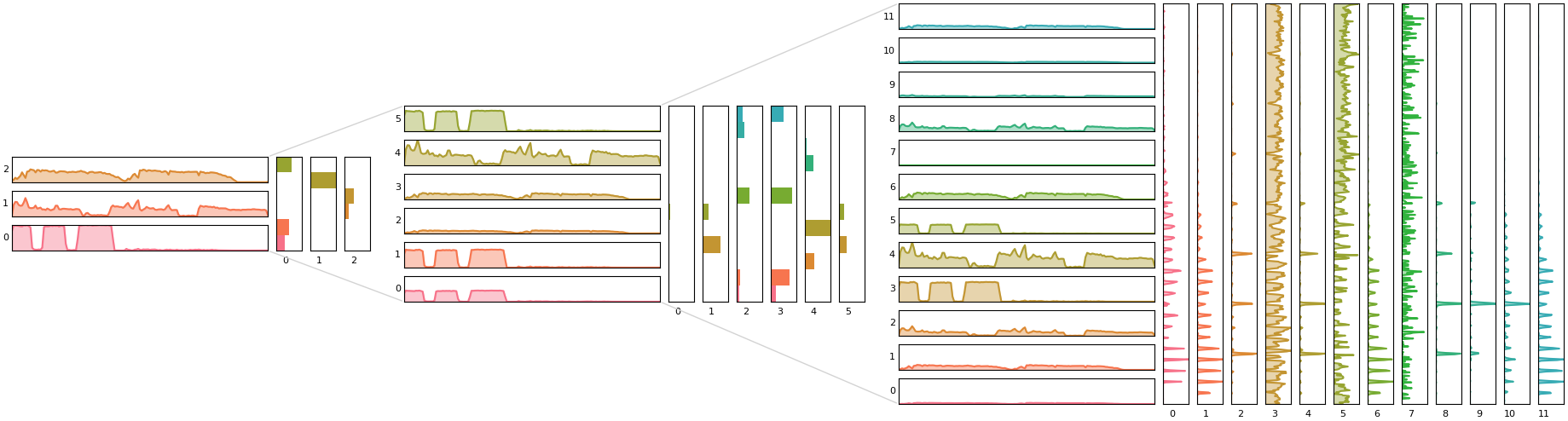}
  \caption{Visualization of a 3-layer sparse DNAE deconstruction of a simple mixture of 3 sources.}
  \label{fig:nae_3_layer}
\end{figure}

Figure \ref{fig:nae_3_layer} contains the visualization of a deconstruction of the described sound into a 3-layer NAE (7 layers in total) with layer sizes set to $L = [3, 6, 12]$, using sparsity regularization. With the increased granularity brought about by the extra level, together with the sparsity constraint, we start observing more diversity among the outer spectra. For instance, it is now possible to observe distinct spectral envelopes for the harmonic spectra of the low-pitch note (numbers 0, 1, 6, and 11). Most notably, the fundamental frequency is missing from spectrum 0. Spectra 3 and 5 share the wideband noise originating from inner activation 0, while the more random-shaped noise spectrum 7 corresponds to a much lower-energy residual noise present throughout the whole sound.  

% model 2025-06-18_22_36_23.572873 resynt 2025-07-08_00_18_20.324352

\subsection{Hierarchical deconstruction}

\begin{figure}[h!]
\center
\subfigure[Activation 1]{\includegraphics[height=2.7cm]{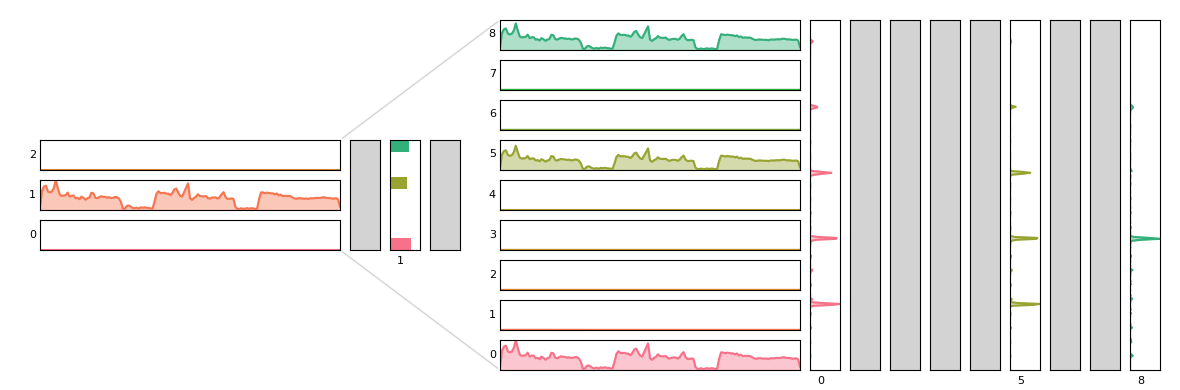}\label{fig:}}
\subfigure[Activation 2]{\includegraphics[height=2.7cm]{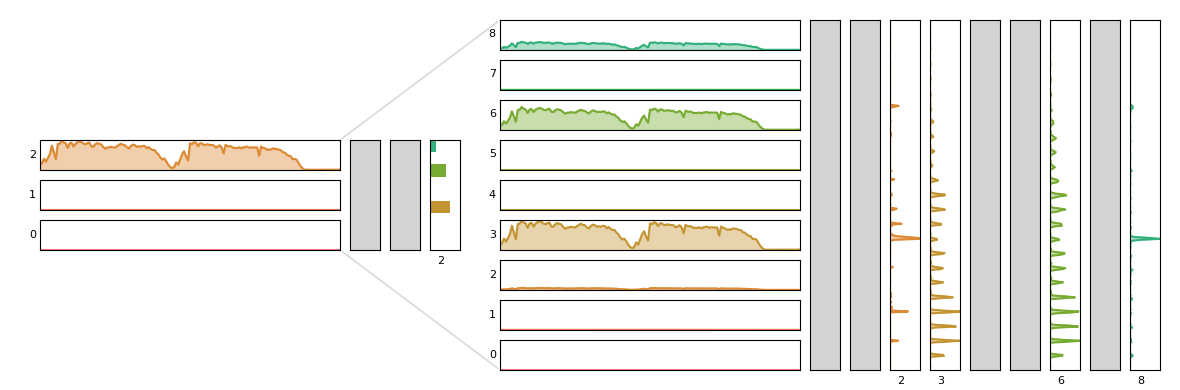}}
\caption{Examples of hierarchical deconstructions of Fig. \ref{fig:nae_2_layer}. Silent weight vectors are greyed out.}
\label{fig:hierarchical}
\end{figure}

The activations of the innermost layer often follow the amplitude profile of elements of a high level of abstraction, such as individual notes, whose contributions are then disseminated among the subsequent layers. We can leverage this to group the outer spectra according to their origin. By selecting one of the inner activations and setting all the other ones to zero (Fig. \ref{fig:hierarchical}), only the spectra that were contributed by that inner activation remain active in the outer layer. This allows us to perform a kind of hierarchical deconstruction, where a sound can be first separated into its notes, and then each note decomposed in increasing levels of granularity as we move towards the outer layers. In a sense, we use the units of the inner layer as clustering indicators of the output components.

\section{Sound manipulation}

So far we have always associated each outer activation $\mf{h}_{(k,L)}$ with its corresponding base $\mf{w}_{(k,L)}$ to create a component for resynthesis, where the $k$ indices are the same. This is always the case in source separation, where the goal is to filter out elements that are contained in the original sound. But in this work, we wish to use the network not only for separation but for sound modification. One way to achieve that is to associate bases and activations that were originally unrelated. If an activation modulates a spectrum to which it was originally unrelated, we are effectively creating new elements that were not present in the original sound. We can distinguish three different cases:
\begin{eqnarray}
	\mathrm{Original\hspace{5pt}component}: & \mf{C}_{(k,l)} & = \mf{w}_{(k,l)} \otimes  \mf{h}_{(k,l)} \\ 
	\mathrm{Cross-component}: & \mf{C}_{(i,j,l)} & = \mf{w}_{(i,l)} \otimes  \mf{h}_{(j,l)}, i \neq j \\
	\mathrm{Cross-layer\hspace{5pt}component}: & \mf{C}_{(i,j,l,m)} & = \mf{w}_{(i,l)} \otimes  \mf{h}_{(j,m)}, l \neq m 
\end{eqnarray}
When creating a non-original component with spectra of the outer mask layer $L$, the resulting Wiener mask (Eq. \ref{eq:wiener}) is no longer a unity-bounded matrix that acts as a filter on the mix. It 
can take values significantly higher than one, filtering out some time-frequency points but enhancing others. To prevent excessively high spectral peaks in the output spectrum, we introduce a bounding factor $\gamma$ to the mask:
\begin{eq}
	\mf{M}_{i,j,l} = \frac{\mf{w}_{(i,L)} \otimes  \mf{h}_{(j,l)} + \frac{\gamma}{K_l}}{\sum_{k=1}^{K_l} \mf{w}_{(k,L)} \otimes  \mf{h}_{(k,l)} + \gamma}. \label{eq:wiener_modif}
\end{eq} 

We can obtain such cross-components either by rearranging the positions of the bases or activations, or by manipulating the inner weights. Changing the original values of the inner weights has the effect of redirecting the energy of the original temporal evolution towards one or several spectra to which it was initially unrelated. We can then create elements that keep the spectral content of the original, but with the temporal structure of another element. For instance, in our toy example, we could have the three square steps modulate one of the harmonic spectra of the trumpet instead of the original white noise. The inner weight matrices can then be thought of as arrays of mixing sliders, and as an extension of the component assignment switchboards of \cite{burred16}, with soft (continuous) rather than binary assignments.  

Instead of individually reassigning or manipulating the components, we can also define a number of random operations in which several assignments are recomputed at once. We propose the following operations:
\begin{itemize}
\item \textbf{Random permutations.} By randomly reassigning whole columns of a given layer's weight matrix to originally unrelated activations of the same layer, we obtain a sound with randomized timbre but similar overall temporal structure, which maintains the same density of sound events as the original sound.
\item \textbf{Weight randomization.} If we freely randomize the values of the weights, besides randomizing timbre, the density of output sound events will almost certainly increase as well. This is because the weight matrices resulting from the original deconstruction tend to be sparse. By replacing them with values sampled from a uniform or normal distribution, without any additional constraint, the original sparse structure will be destroyed. This results in a greater spread of energy across the units of all subsequent layers, and ultimately much denser sounds. It is possible to apply this operation selectively to only given columns of the weight matrices to control the degree of timbre randomization and output density. For instance, randomizing the first weight column of the first layer of Fig. \ref{fig:nae_3_layer} would randomize the timbre associated with the three square bursts while keeping the original trumpet notes intact.
\item \textbf{Event-internal randomization.} To obtain a more subtle timbre randomization that keeps sparsity, instead of replacing the weights with completely new randomly sampled values, we can multiply the random values with the original ones. If the range of the multiplicative sampled values is moderate (for instance ,by sampling from a narrow uniform distribution), the sparsity structure and overall density of the timbre-randomized output sound will barely be affected. As with the previous one, this operation can also be applied to a given subset of weight columns.
\end{itemize}
All these operations can be defined to be operated on a given decoder layer, and thus with a controllable level of granularity. We refer to the companion website referenced above for a selection of sound examples.

\section{Conclusions}

In this work, we introduced Non-Negative Autoencoders (NAEs) as a novel and interpretable neural framework for sound deconstruction, analysis, and creative manipulation. Building upon the structure and constraints of Non-Negative Matrix Factorization (NMF), we extended the concept to deep architectures, enabling hierarchical decompositions with adjustable levels of granularity and abstraction.

By enforcing non-negativity through projected gradient descent and carefully selecting the optimization strategy and loss function, we achieved stable training and interpretable internal representations even for deeper networks. Our experiments demonstrated that multi-layer NAEs can extract meaningful temporal and spectral components from complex sounds and organize them in a structured, layered fashion. These representations not only allow for detailed analysis and resynthesis but also enable expressive and controllable manipulations, such as timbre randomization, event recombination, and cross-component synthesis. %This aligns with the needs of interactive and exploratory sound design, placing the user at the center of the creative process.

An observed limitation is that the system becomes increasingly difficult to parameterize and optimize as the number of layers increases. More structured regularization techniques, together with more thorough convergence analysis, will likely be needed to ensure robustness and maintain diversity of the extracted components in deeper architectures.

In future work, we also plan to implement extensions to stereo and multi-channel audio. We also aim to investigate alternative constraints and regularization strategies to enhance component diversity and control across more complex and varied sounds. As a longer-term goal, we will investigate the relevance and feasibility for the task at hand of probabilistic reformulations via Non-negative Variational Autoencoders (NVAEs) \cite{xie2023}.

\section{Acknowledgements}

The first author gratefully acknowledges Carmine Emanuele Cella for the invitation to spend a semester as a visiting scholar at CNMAT, which made this collaboration possible. Thanks also to Edmund Campion, Andrew Blanton, Luke Dzwonczyk, Jeremy Hunt, and Jeremy Wagner for their help and fruitful discussions.

\bibliographystyle{ieeetr}  
\bibliography{jj_cnmat}

\end{document}